\documentclass[prb,aps, twocolumn, amsmath,amssymb,superscriptaddress]{revtex4}
\usepackage{braket}
\usepackage{adjustbox}
\usepackage{float}
\usepackage{amsmath}
\usepackage{amssymb}
\usepackage{amsfonts}
\usepackage{euscript}
\usepackage{enumerate}
\usepackage{hhline}
\usepackage{pslatex}
\usepackage{tabularx}
\usepackage[usenames,dvipsnames]{xcolor}
\usepackage{threeparttable}
\usepackage{graphicx}
\usepackage{dcolumn}
\usepackage{bm}
\usepackage[sort&compress]{natbib}
\usepackage{setspace}

\makeatletter

\renewcommand{\@biblabel}[1]{#1. }
\renewcommand{\@dotsep}{500}
\renewcommand{\@pnumwidth}{0em}
\renewcommand{\l@figure}[2]{
\@dottedtocline{1}{1.5em}{2em}{Figure #1}{}\vspace{15pt}}
\begin{document}
\begin{spacing}{1.25}
\title{A highly efficient integrated source of twisted single-photons}

\author{Bo Chen}
\affiliation{State Key Laboratory of Optoelectronic Materials and Technologies, School of Physics, Sun Yat-sen University, Guangzhou 510275, China}

\author{Yuming Wei}
\affiliation{State Key Laboratory of Optoelectronic Materials and Technologies, School of Physics, Sun Yat-sen University, Guangzhou 510275, China}

\author{Tianming Zhao}
\affiliation{State Key Laboratory of Optoelectronic Materials and Technologies, School of Physics, Sun Yat-sen University, Guangzhou 510275, China}

\author{Shunfa Liu}
\affiliation{State Key Laboratory of Optoelectronic Materials and Technologies, School of Physics, Sun Yat-sen University, Guangzhou 510275, China}

\author{Rongbin Su}
\affiliation{State Key Laboratory of Optoelectronic Materials and Technologies, School of Physics, Sun Yat-sen University, Guangzhou 510275, China}

\author{Beimeng Yao}
\affiliation{State Key Laboratory of Optoelectronic Materials and Technologies, School of Physics, Sun Yat-sen University, Guangzhou 510275, China}

\author{Ying Yu}
\affiliation{State Key Laboratory of Optoelectronic Materials and Technologies, School of Electronics and Information Technology, Sun Yat-sen University, Guangzhou 510275, China}

\author{Jin Liu}
\thanks{http://spe.sysu.edu.cn/photon/}
\affiliation{State Key Laboratory of Optoelectronic Materials and Technologies, School of Physics, Sun Yat-sen University, Guangzhou 510275, China}

\author{Xuehua Wang}
\affiliation{State Key Laboratory of Optoelectronic Materials and Technologies, School of Physics, Sun Yat-sen University, Guangzhou 510275, China}
\date{\today}

\begin{abstract}
	Photons with a helical phase front (twisted photons) can carry a discrete, in principle,	unbounded amount of orbital angular momentum (OAM). Twisted single-photons have been demonstrated as a high-dimensional quantum system with information processing ability far beyond the widely used two-level qubits. To date, the generations of single-photons carrying OAM merely rely on the non-linear process in bulk crystals, e.g., spontaneous parametric down-conversion (SPDC), which unavoidably limits both the efficiency and the scalability of the source. Therefore, an on-demand OAM quantum light source on a semiconductor chip is yet illusive and highly desirable for integrated photonic quantum technologies. Here we demonstrate highly-efficient emission of twisted single-photons from solid-state quantum emitters embedded in a microring with angular gratings. The cavity QED effect allows the generations of single-photons and encoding OAM in the same nanostructure and therefore enables the realization of devices with very small footprints and great scalability. The OAM states of singe-photons are clearly identified via quantum interference of single-photons with themselves. Our device may boost the development of integrated quantum photonic devices with potential applications towards high-dimensional quantum information processing.

\end{abstract}

\maketitle

 A unique degree of freedom associated with photons is the orbital angular momentum (OAM) which appears in a discrete step of $l$$\hbar$ \cite{Allen1992}, where $l$ is an unbounded integer and $\hbar$ is the Planck constant. The discovery of twisted photons has led to a plethora of enchanting applications including micro-manipulation\cite{Grier2003,Padgett2011}, optical microscopy\cite{Abouraddy2006}, optical communication\cite{WangJ2012,Vallone2014,Bozinovic2013} and quantum information processing\cite{Mair2001, Barreiro2010, Dambrosio2012, Parigi2015,Tomer2018}. In particular, the infinite optical states lying in OAM significantly boost the capacities of optical communication and information processing in both classic and quantum regimes. In the non-classic domain, twisted single-photons have been employed as a high-dimensional quantum system to push the quantum information processing to a level that the conventional two-level qubits are not able to reach. E.g., single-photons carrying OAM have facilitated the experimental realizations of enhanced robustness against eavesdropping and quantum cloning \cite{BechmannPH2000,CerfNJ2002}, high-dimensional quantum entanglements \cite{RomeroJ2012,ChenL2017}, long-distance high-dimensional quantum key distribution \cite{KrennM2015,KrennM2016,SitA2017} and quantum teleportation of multiple-degrees of freedom of a single-photon \cite{WangXL2015}.

\begin{figure*}[htpb]
	\begin{center}
		\includegraphics[width=0.9\linewidth]{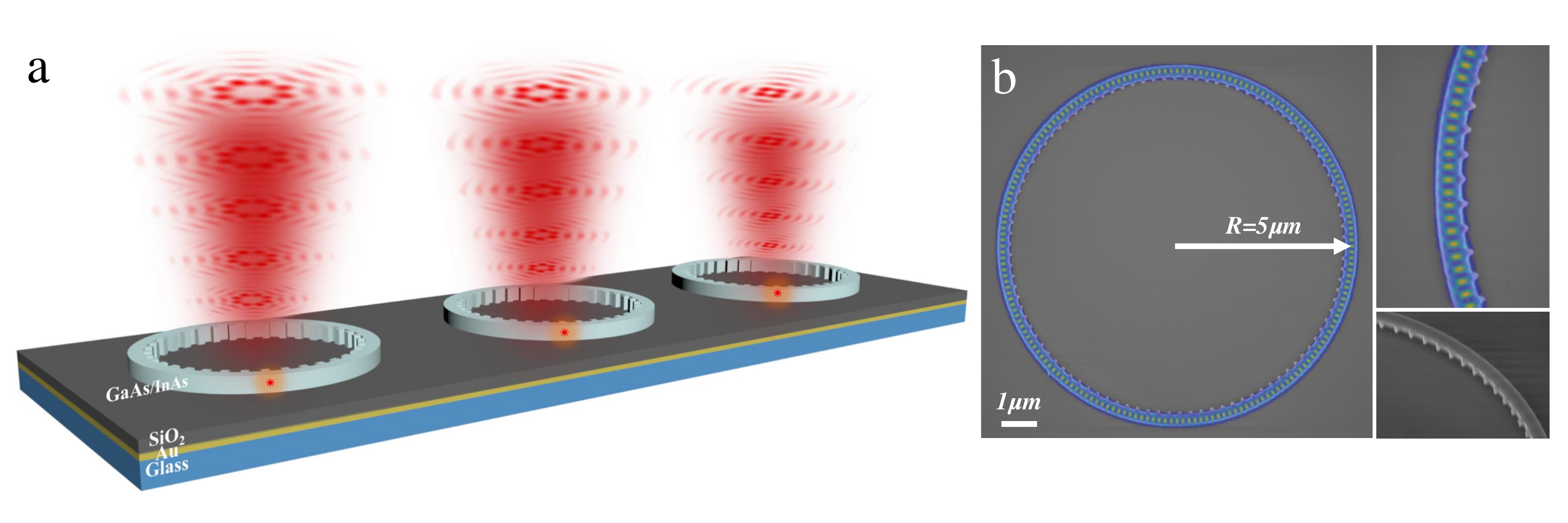}
		\caption{
			\textbf{Microring resonators embedded with QDs for the generation of single-photons carrying the quantum superposition states of OAM.}
			(a) Schematic of the devices emitting single-photons carrying OAMs. (b) SEM images of a fabricated microring resonator ($R = 5~\mu m,q=98$) with the field intensity distribution superimposed.		
		}
		\label{fig:Fig1}
	\end{center}
\end{figure*}

However, the current developments of OAM single-photon sources face formidable challenges in both efficiency and scalability, preventing the high-dimensional quantum information processing moving from the proof-of-concept stage to the practical applications. Hitherto, all the existing single-photons carrying OAM have been realized by generating single-photons with spontaneous parametric down-conversion (SPDC) process in bulk nonlinear crystals and subsequently coupling the emitted single-photons to a mode transformer to encode OAM. The SPDC is intrinsically probabilistic\cite{Mair2001,Tomer2018,WangXL2018} and greatly suffers from the trade-off between the brightness and single-photon purity therefore usually operate at a low pump power with an efficiency of only a few percents. In addition, encoding the SPDC generated single-photons with OAMs requires the introduction of an extra mode transformer, e.g., a spatial light modulator, and therefore inevitably adds extra loss and enlarges the device footprints.

Epitaxial quantum dots (QDs) may serve as a very effective solution for dramatically improving the efficiency and the scalability of the quantum OAM sources. They have been demonstrated as deterministic single-photon sources simultaneously exhibiting high degrees of brightness, single-photon purity and indistinguishability \cite{DingX2016, Unsleber2016, Senellart2017, HeYM2017}, which facilitates the advance towards quantum supremacy with single-photons \cite{WangHBoson20}. Epitaxial QDs, e.g., InAs/GaAs QDs, are also highly compatible with the modern semiconductor fabrication process, allowing fabrications of a large number of light emitting devices with very compact footprints on a semiconductor chip.

In this work, we experimentally realize the first integrated sources of twisted single-photons, exhibiting high-efficiency, potential scalability and compact footprint. The schematic of our devices is shown in Fig.~1(a). Single QDs are embedded in microrings \cite{Cai2012, Strain2014, MiaoP2016, ZhangJuan2018} that sit on a thin $\mathrm{SiO_{2}}$ layer with a backside metallic layer. The emitted single-photons from QDs are coupled to the standing wave formed by the interference of the two counter-propagating ( the clockwise (CW) mode and counterclockwise (CCW) mode) whispering gallery modes (WGMs) of the microrings and subsequently scattered by the angular gratings to the free-space, forming upwards going single-photons carrying OAM. The downwards emitted single-photons are efficiently reflected upwards by the broadband  metallic mirror on the bottom. We present the details of epitaxial growth and device fabrication in supplementary information (SI). The scanning electron beam (SEM) images of the fabricated device are presented in Fig.~1(b) in which the angular gratings on the inner sidewall of the microring can be identified. By periodically modulating the refractive index in the azimuthal direction, the WGMs that confined in the microring resonator are scattered out to the free-space. For a device with fixed number of gratings $q$, the resonant WGMs order $p$ determines the topological charge $l$ of the emitted optical vortex, as $l=\mathrm{sign}(p)(p-q)$ \cite{Cai2012}. Therefore the counter-propagating WGMs at the same resonant wavelength have exactly inversed phase gradient, which introduces spiral photons with equal but opposite topological charges. We superimpose the standing wave pattern formed by the interference of the two counter-propagating WGMs on the SEM images of the real device, as shown in Fig.~1(b). 

\begin{figure*}[htpb]
	\begin{center}
		\includegraphics[width=0.7\linewidth]{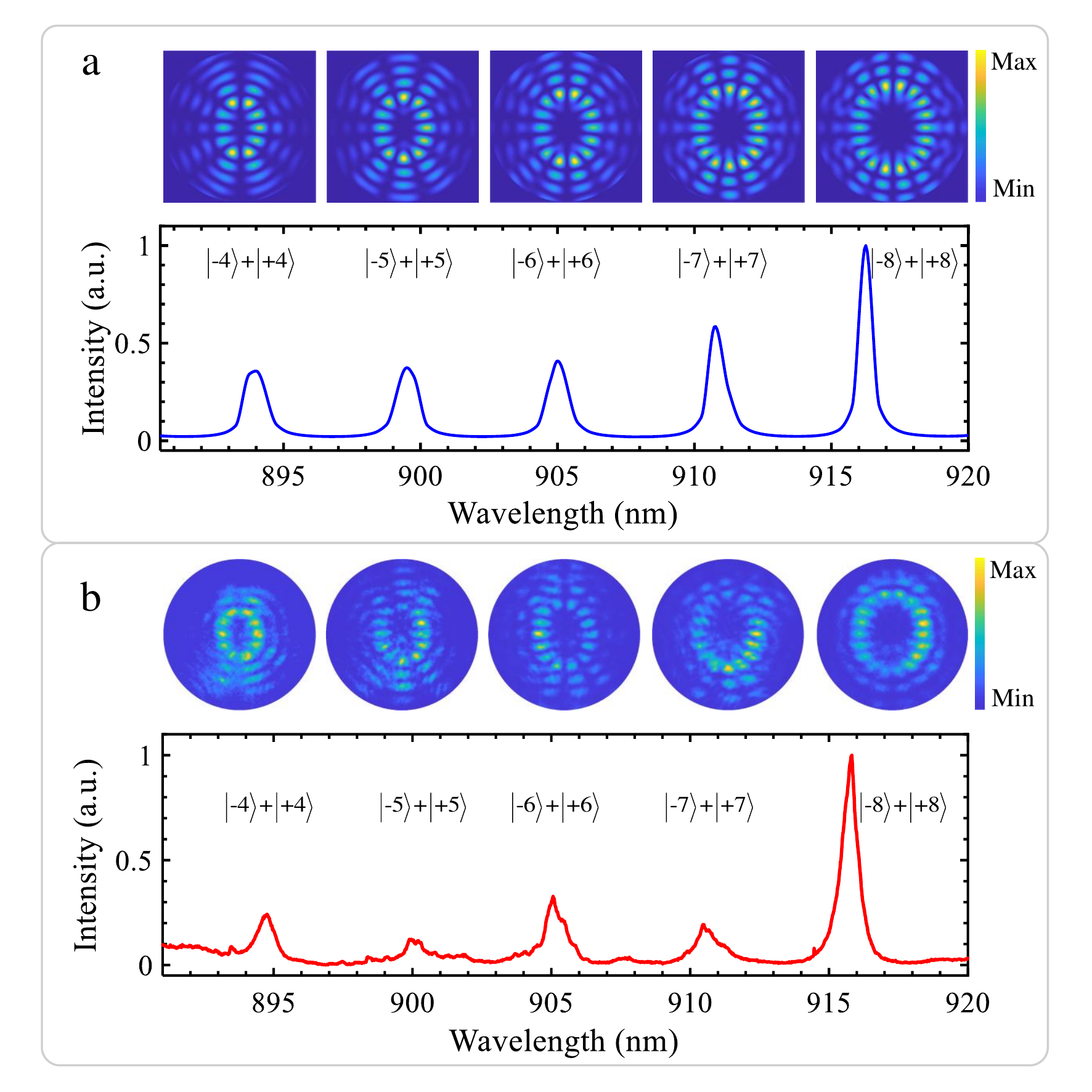}
		\caption{\textbf{Characteristics of the microring cavity modes carrying OAMs.} (a) Simulated radiation spectrum (blue line) of the microring resonator with $R = 5~\mu m, q=98$ and the corresponding far-field patterns projected to horizontal polarization. (b) Measured radiation spectrum (red line) and the far-field patterns for the device simulated in (a).}
		\label{fig:Fig2}
	\end{center}
\end{figure*}

The OAM state of the photons scattered out by angular gratings is written as $|\Phi\rangle_l = \frac{1}{\sqrt{2}}[|+l\rangle+\mathrm{exp}(i\delta(\varphi)|-l\rangle)]$ \cite{Fickler2012}. The topological charges of such a superposition state can be directly extracted from the far-field patterns formed by the interference of the emitted single-photons with themselves. Due to the nature of the cylindrical vector vortices \cite{ZhuOL2013,ZhuOL2014}, the polarization state of the emitted photons for different cavity modes can be radially polarized (RP), azimuthally polarized (AP) or hybrid RP/AP. The topological charges of hybrid RP/AP vortex states (e.g., the states in Fig.~2) can be identified by counting the number of the antinodes, $n=2(|l|+1)$, in the far-field pattern when projected to the linear polarization, see the detailed discussion in the SI. Fig.~2(a) shows the finite difference time domain (FDTD) simulated radiation spectrum of the cavity modes carrying OAMs and their corresponding far-field patterns projected to horizontal polarization for the fabricated device in Fig.~1(b). 

In the experiment, the cavity modes of our OAM devices can be efficiently probed by optically exciting the QDs with a high laser power of $\sim$105~$\mu$W. At this power, the QD ensemble serves as a broadband classic light source for lighting up the cavity modes. The spanning wavelength ranges from 890~nm to 920~nm due to the power broadening. The Q-factors of the cavity modes are of $\sim$1000 mostly limited by the grating scattering. In the meantime, the far-field patterns of the cavity modes can be steadily accessed by spectrally filtering each cavity resonances with a tunable band-pass filter and imaging the upwards emitted photons in the back focal plane of the confocal microscope with an electron multiplying charge-coupled device (EMCCD), see Fig.~2(b) and Fig.~S4 in SI. Both the radiation spectrum and the far-field patterns obtained from the experiments agree well with the FDTD simulations (see the details of the numerical simulations in SI.) in Fig.~2(a), which validates the generation and identification of the superposition states of OAMs in our devices. 

\begin{figure*}[htpb]
	\begin{center}
		\includegraphics[width=1.0\linewidth]{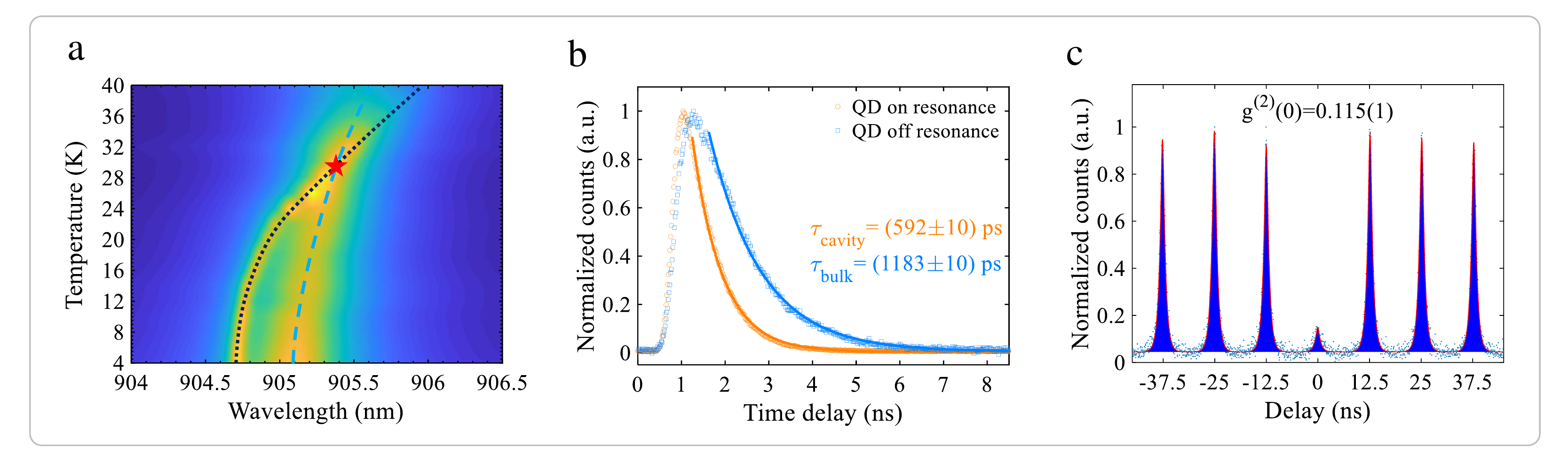}
		\caption{\textbf{Purcell-enhanced single-photons carrying OAMs.}
			(a)  Spectra of the coupled cavity-QD system as a function of the temperature varying from  4~K to 40~K. The dot dashed line indicates the QD and the long dashed line denotes the cavity mode. The cross point is about at 30~K marked by the red star. (b) Time-resolved measurements for the QDs on and off the resonance, revealing a pronounced reduction of the radiative lifetime.  (c) Auto-correlation histogram measured under pulsed excitation with  a Hanbury-Brown and Twiss interferometer. The second-order correlation $g^{(2)}(0)=0.115(1)$ is calculated from the integrated area in the zero delay peak divided by that of peaks away from zero delay.}
		\label{fig:Fig3}
	\end{center}
\end{figure*}

The isolated single QD emission (dotted line in Fig.~3(a)) can be clearly observed when lowering down the excitation power to $\sim$350~nW, which is close to the saturation power of single QDs. At 4~K, a single QD at 904.72~nm is spectrally near to the cavity mode (dashed line in Fig.~3(a)) with the topological charge of $|l|=6$ (see the details of identification of the topological charge $l$ in Fig.~S4 of the SI) at 905.11 nm. To maximally funnel the emitted single-photons from the QD to the cavity mode carrying OAM, we change the temperature of the sample to tune the single QD line into the cavity resonance. Due to the temperature dependences of the bandgap and the refractive index of GaAs \cite{Blakemore1982,GehrsitzS2000}, both QD emission and cavity mode shift to the longer wavelength as the increase of the temperature. As the QD line red-shifts faster than the cavity mode, we obtain an on-resonance condition at 30~K. The lifetime of the QD on the resonance condition is significantly shorter than that of the QD off the cavity resonance, which unequivocally reveals the efficient coupling between the QD and the cavity mode, as shown in Fig.~3(b). A Purcell factor of $\sim$2 is extracted, which results in a spontaneous emission factor, i.e., $\beta$ factor of 0.67 by using $\beta=F_{p}/(F_{p}+1)$. Therefore, $67\%$ of the emitting single-photons are coupled to the cavity mode and consequently carry the OAM. We note that the simulated Purcell factor is $\sim$7, which gives rise to a $\beta$ factor close to 90\%. The deviation of the measured Purcell factor from the simulated number could be attributed to the fact that the QD is not exactly located at the maximal intensity of the cavity mode. The single-photon purity of the emitted photons carrying OAM is clearly identified from the second-order coherence measurement shown in Fig.~3(c), in which nearly vanished multi-photon probability is observed at zero delay [$g^{(2)}(0)=0.115(1)$].

Finally, we characterize the OAM states of emitted single-photons via imaging the interference patterns of single-photons with themselves. Fig.~4(a) shows the spectrum for the single QD line that maximally coupled to the cavity mode at 30~K. The single QD line (shown in blue) at~905.11~nm is spectrally filtered by a narrow band-pass filter and sent into an EMCCD for imaging. The ``near-field" (image of emission from the sample surface in the experiment and calculated intensity distribution for $\lambda/2$ above the sample surface in the simulation) and far-field intensity distributions for the single-photons carrying OAM superposition states with $|l\rangle =\mathrm{|-6\rangle+|+6\rangle}$ are presented on the right side of Fig.~4(b,c) respectively, which agrees well with the numerical simulations on the left side of Fig.~4(b,c). Analogous to Young's double slits interference experiment with single-photons, we observe the transition of the far-field intensity distribution from randomly detected photons to an interference pattern with multiple lobes on the EMCCD when increasing the image acquisition time to several tens of seconds. Such interference patterns arise from the superposition state of two equal and opposite OAM modes of single-photons \cite{Fickler2012, KrennM2015}. We then project the single-photons to different linear polarizations with a polarizer. As illustrated at the lower side of Fig.~4(d), the projected far-field patterns exhibit the feature of hybrid RP/AP cylindrical vector vortices showing characteristic orientations at different polarizations, which agrees excellently well with our numerical simulations at the upper side of Fig.~4(d). As mentioned previously, the topological charges of the OAM states can be directly recognized by counting the number of antinodes in the far-field patterns of the emitted single-photons. The systematic and quantitative agreements between the experiments and simulations for the intensity distributions of single-photons in both ``near-field" and far-field unequivocally reveals the OAM carried by the emitted single-photons.

\begin{figure*}[htpb]
	\begin{center}
		\includegraphics[width=0.8\linewidth]{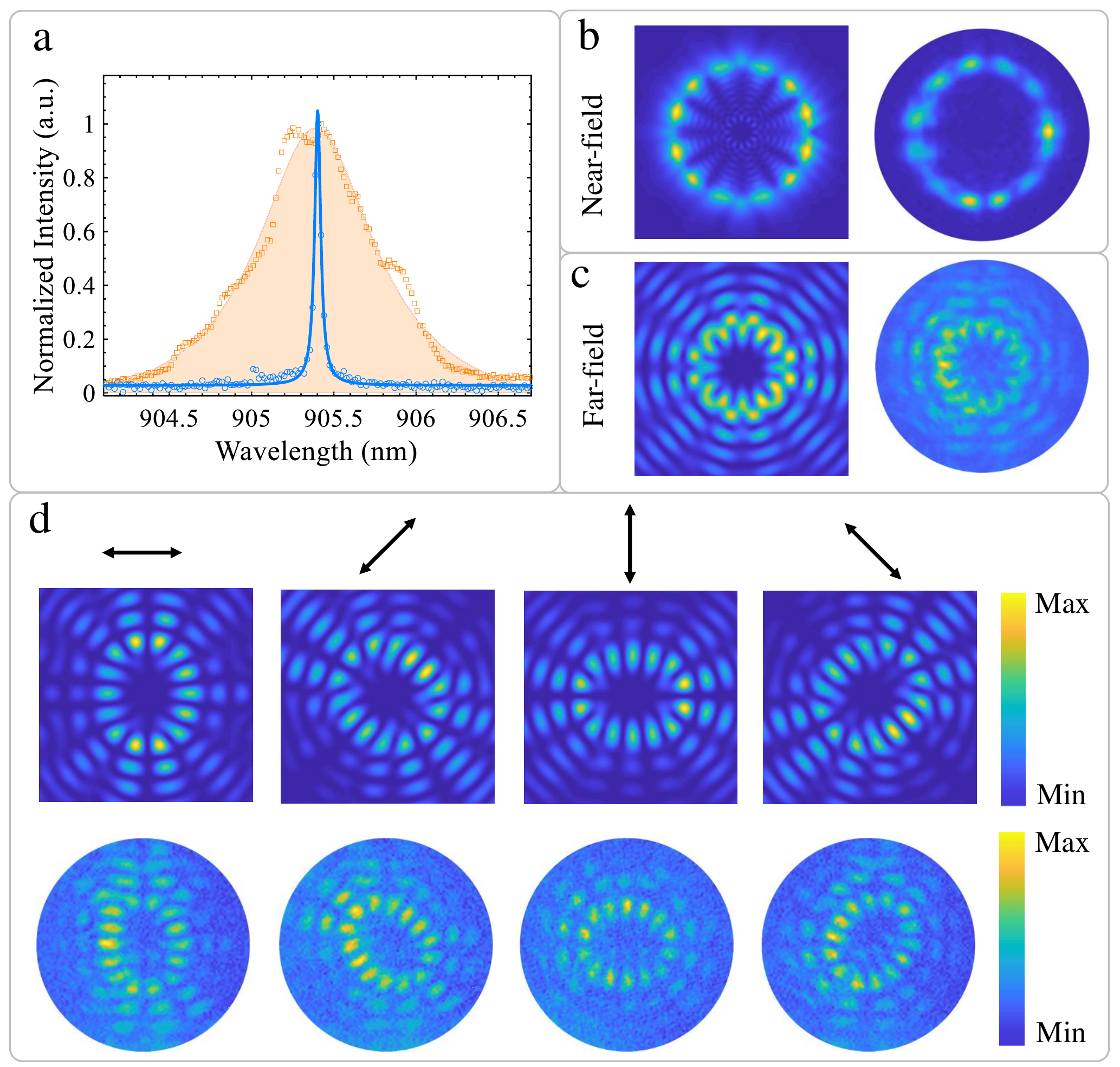}
		\caption{\textbf{ Near-field and far-field distribution of single-photons carrying the quantum superposition state of OAMs.}  (a) PL spectrum for the single-photon emission (blue line) coupled to the cavity mode. The cavity mode with $\mathrm{|l\rangle=|-6\rangle+|+6\rangle}$ measured at a high power is denoted with the orange area. (b) The simulated (left) and measured (right) near-field intensity distributions for the emission from the QD located in the microring. (c) The far-field intensity distributions corresponding to (b). (d) Upper row: the 3D-FDTD simulations of the intensity distributions for the far-field projected to linear polarization along horizontal, diagonal, vertical, anti-diagonal directions; lower row: the corresponding measured far-field intensity distributions projected to linear polarization along differnt directions for the emitted single-photons carrying OAM. The directions of the polarization projection are indicated by the arrows.}
		\label{fig:Fig4}
	\end{center}
\end{figure*}	


In summary, we have proposed and experimentally demonstrated a highly-efficient OAM single-photon source from QDs embedded in a fully integrated microring cavity with angular gratings. The OAM single-photons are realized by coupling spontaneous emission from QDs to the standing wave formed by the interference of the CW and CCW WGMs of the microring. The single-photon carrying OAMs are subsequently extracted to the free-space via periodic modulation of the refractive index along the azimuthal direction with gratings in the inner sidewall of the microring. We directly identified the topological charges of the OAM states by imaging the far-field patterns of the emitted single-photons without an external interference scheme for characterizing the wavefront structure. The efficiency and the footprint of the device are  benefit from the Purcell effect that both enhances the emission rate of QDs and encodes the OAM to the single-photons. We believe that the topological charges of OAM single-photons in our system could also be accurately controlled by employing the recently demonstrated strain-tuning technique on QDs \cite{YuanX2018}. Moving forward, more advanced solid-state OAM quantum light sources could be envisioned by further exploring the potential of the integrated QD platform. E.g., fabricating spiral phase plates with a focused ion beam (FIB) on the top surface of micro-pillar cavities could lead to bright and indistinguishable single-photons carrying OAMs with single topological charges. The hyper-entanglement states \cite{Prilmuller2018} can also be generated by coupling the entangled photons emitted by QDs to the specially designed photonic modes carrying OAM. With further technological improvements, our results may immediately boost the developments of a few long-sought high-dimensional quantum information processing experiments including quantum communication with photons in spatial modes, quantum teleportation and entanglement swapping of high-dimensional quantum states etc.


\end{spacing}
\end{document}